\documentclass[conference,a4paper]{IEEEtran}
\usepackage{amsmath,amssymb}
\ifCLASSINFOpdf
  \usepackage[pdftex]{graphicx}
\else
  \usepackage[dvips]{graphicx}
\fi
\usepackage{balance}
\usepackage{url}
\graphicspath{{figs/}}   
\usepackage{booktabs}
\usepackage{tikz}
\usetikzlibrary{positioning,calc}

\begin{document}

\onecolumn
\thispagestyle{empty}
\vspace*{\fill}
\begin{center}
	{\large This work has been submitted to the IEEE for possible publication.
		Copyright may be transferred without notice, after which this version may no
		longer be accessible.\par}
	\vspace{1.5em}
	Submitted to 21st European Conference on Antennas and Propagation (EuCAP 2027).
\end{center}
\vspace*{\fill}
\clearpage
\twocolumn

\title{Opportunities for Direct-to-Cell Polarization MIMO in the
Downlink of a LEO Satellite Footprint}

\author{\IEEEauthorblockN{Rainer Bachl\IEEEauthorrefmark{1}, Matteo Ferro\IEEEauthorrefmark{1}, Bingguang Peng\IEEEauthorrefmark{2}, Muhammad Nabeel\IEEEauthorrefmark{1}}
\IEEEauthorblockA{\IEEEauthorrefmark{1}Huawei Heisenberg Research Center, Munich, Germany,
\{rainer.bachl, matteo.ferro, muhammad.nabeel8\}@huawei.com}
\IEEEauthorblockA{\IEEEauthorrefmark{2}Huawei Shanghai Research Institute, Shanghai, China, pengbingguang@huawei.com}
}

\maketitle
\setlength{\abovedisplayskip}{4pt plus 1pt minus 2pt}
\setlength{\belowdisplayskip}{4pt plus 1pt minus 2pt}
\setlength{\abovedisplayshortskip}{1pt plus 1pt}
\setlength{\belowdisplayshortskip}{2pt plus 1pt minus 1pt}

\begin{abstract}
A single low Earth orbit (LEO) satellite offers almost no spatial degrees of freedom, but its two polarizations can carry a second layer to the handset. The downlink spectral efficiency is mapped over the footprint of a satellite at 330\,km altitude above Munich. The channel is a cascade of Jones factors including the phased-array payload, Faraday rotation, a dual-polarized land mobile satellite channel and a standardized handset antenna element at three attitudes. For a handset with boresight to zenith and 6\,dB SNR, ideal precoding selects rank~2 with probability of at least 0.95 within 340\,km of the nadir point. With closed-loop precoding, such a handset on the nadir track is in rank~2 during 99\,\% of a pass. Upright handsets also select rank~2 wherever they face the satellite, and rank selection improves the mean spectral efficiency over the footprint by up to 49\,\%.

\end{abstract}

\begin{IEEEkeywords}
	NR~NTN, direct-to-cell, handset attitude, polarization MIMO, rank selection, land mobile satellite channel.
\end{IEEEkeywords}

\section{Introduction}
\label{sec:intro}
Direct-to-cell connectivity, in which handsets communicate directly
with satellites in low Earth orbit (LEO), is part of 5G New Radio
(NR) through its non-terrestrial network extensions, NR~NTN~\cite{tr38821}. On a link between a single satellite
and a handset, multiple-input multiple-output (MIMO) transmission has
almost no spatial degrees of freedom. The antennas at both ends are
small compared with their distance, and the line of sight (LOS)
dominates. The two
orthogonal polarizations remain as the only practical MIMO resource of
such a link~\cite{arapoglou2011review}. Measurements of the
dual-polarized land mobile satellite (LMS)
channel~\cite{kingthesis2007,nikolaidis2017} indicate that the polarizations
can carry separate streams, and polarization multiplexing has been
demonstrated for satellite
broadcasting~\cite{byman2016}. However, NR~NTN
uses polarization to separate beams under frequency
reuse~\cite{tr38821} and signals a single downlink polarization per
cell~\cite[Sec.~6.3.2]{ts38331}, without polarization multiplexing.
Recent surveys of MIMO over
satellite consider MIMO across several satellites for direct-to-device
links~\cite{bakhsh2025} or treat the polarizations as a resource for
index modulation~\cite{heo2023}. Whether a single satellite can provide a second polarization layer to a handset is therefore an open question. Four
effects act on the two polarizations during a pass. The port states of the payload lose orthogonality as the beam is
steered away from nadir, and the ionosphere rotates the plane of
polarization. The environment couples the two polarizations, and the
attitude of the handset turns its ports and scales their gain. All four depend on the position of the handset
relative to the satellite, so the benefit of a rank-2 transmission,
i.e., of two layers, varies across the satellite footprint.

To determine where a second polarization layer is beneficial, we model
the downlink from an S-band LEO satellite to a handset as a cascade of
the four effects above. The cascade combines the scan geometry of the
payload, Faraday rotation in the ionosphere, a dual-polarized LMS channel, and a
standardized handset antenna element under different attitudes. With this model, we evaluate rank selection with ideal and
with closed-loop precoding. Over the satellite footprint, we obtain the
areas in which rank~2 is selected, the spectral efficiency (SE) gain of
rank selection over rank~1, and the percentage of a pass in rank~2.

\section{System Model}
\label{sec:model}
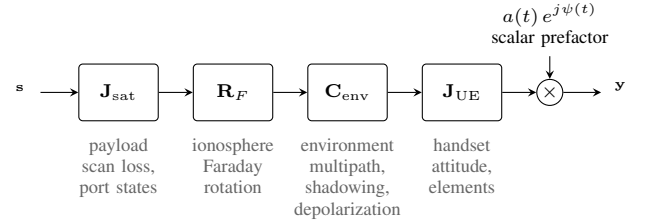
\begin{figure}[t]
	\centering
	\begin{tikzpicture}[
		font=\scriptsize,
		blk/.style={draw, rounded corners=1.5pt, minimum height=7.5mm,
			minimum width=10.5mm, align=center, inner sep=1.5pt},
		own/.style={align=center, text=black!60, font=\scriptsize},
		>=stealth]
		\node[blk] (jsat) {$\mathbf J_{\mathrm{sat}}$};
		\node[blk, right=4.5mm of jsat] (rf)   {$\mathbf R_F$};
		\node[blk, right=4.5mm of rf]   (cenv) {$\mathbf C_{\mathrm{env}}$};
		\node[blk, right=4.5mm of cenv] (jue)  {$\mathbf J_{\mathrm{UE}}$};
		\node[draw, circle, right=4.5mm of jue, inner sep=0.6pt] (mul) {$\times$};
		\draw[->] ($(jsat.west)+(-5mm,0)$) -- (jsat.west)
		node[pos=0, above right=-1mm and -4.5mm, font=\tiny] {$\mathbf{s}$};
		\draw[->] (jsat) -- (rf);
		\draw[->] (rf) -- (cenv);
		\draw[->] (cenv) -- (jue);
		\draw[->] (jue) -- (mul);
		\draw[->] (mul.east) -- ++(5mm,0)
		node[above left=-1mm and -4.5mm, font=\tiny] {$\mathbf{y}$};
		\node[above=3mm of mul, align=center, font=\scriptsize] (sc)
		{$a(t)\,e^{j\psi(t)}$\\scalar prefactor};
		\draw[->] (sc) -- (mul);
		\node[own, below=1mm of jsat] {payload\\scan loss,\\port states};
		\node[own, below=1mm of rf]   {ionosphere\\Faraday\\rotation};
		\node[own, below=1mm of cenv] {environment\\multipath,\\shadowing,\\depolarization};
		\node[own, below=1mm of jue]  {handset\\attitude,\\elements};
	\end{tikzpicture}
	\caption{Channel model~\eqref{eq:cascade} as a cascade of Jones factors
		from the payload ports $\mathbf{s}$ to the handset ports $\mathbf{y}$,
		with the effects represented by each factor.}
	\label{fig:cascade}
\end{figure}

We model the channel between the two polarization ports of the satellite
payload and the two receive ports of the handset at time $t$ as a
narrowband cascade of $2\times2$ Jones factors~\cite{jones1941},
\begin{equation}
	\mathbf H(t)=a(t)\,e^{j\psi(t)}\,
	\mathbf J_{\mathrm{UE}}(t)\,\mathbf C_{\mathrm{env}}(t)\,
	\mathbf R_F(t)\,\mathbf J_{\mathrm{sat}}(t), \label{eq:cascade}
\end{equation}
as illustrated in Fig.~\ref{fig:cascade}. Here, $\mathbf J_{\mathrm{sat}}$
is the payload factor, $\mathbf R_F$ the Faraday rotation,
$\mathbf C_{\mathrm{env}}$ the environment factor, and
$\mathbf J_{\mathrm{UE}}$ the factor of the handset, or user equipment
(UE). The scalar prefactor is common to all entries of $\mathbf H$. Its
amplitude $a(t)$ contains only the ionospheric scintillation, since the
path gain enters through the signal-to-noise ratio (SNR) in
Section~\ref{sec:la}. Its phase $\psi(t)=-2\pi f_c\tau_0(t)$ at the
carrier frequency $f_c$ follows from the bulk delay $\tau_0(t)=d(t)/c$
over the slant range $d(t)$, with $c$ the speed of light. The rate
$\dot\psi/2\pi$ of the phase is the Doppler shift of the link. Both are assumed to be compensated at the handset from the satellite
ephemeris. The prefactor thus sets the SNR but cancels from the
condition number $\kappa=\sigma_1/\sigma_2\ge1$, with
$\sigma_1\ge\sigma_2$ the singular values of $\mathbf H$, i.e., the
gains of its two modes. The condition number measures how much weaker
the second mode is than the first.

\subsection{Payload Factor $\mathbf J_{\mathrm{sat}}$}
\label{sec:jsat}
The payload is a planar array of co-located crossed elements, aligned
with the velocity and the orbit normal, whose beam is steered toward the
handset. Its factor is
\begin{equation}
	\mathbf J_{\mathrm{sat}}=g\,\mathbf T,\qquad
	g=(\cos\alpha\cos\beta)^{P/2},
	\label{eq:jsat}
\end{equation}
where $g$ is the amplitude scan loss, $\alpha$ and $\beta$ are the
angles of the beam from the array normal, measured in the two principal
planes, which contain the velocity axis and the orbit normal,
respectively, and $P$ is the scan-loss exponent. We set $P=1.3$, a
roll-off steeper than the $\cos\theta$ of an ideal matched
array~\cite{hannan1964}, to account for the embedded element pattern and
implementation losses of practical arrays. The port
states $\mathbf T$ in~\eqref{eq:jsat} follow from the projection
$\mathbf A$ of the two element axes onto the plane transverse to the
direction to the handset,
\begin{equation}
	\mathbf A=\begin{bmatrix}\cos\theta\cos\phi & \cos\theta\sin\phi\\
		-\sin\phi & \cos\phi\end{bmatrix}
	=\mathrm{diag}(\cos\theta,1)\,\mathbf R(\phi),
	\label{eq:jsat_proj}
\end{equation}
where $\theta$ is the scan angle from the array normal, $\phi$ the
azimuth of the beam in the array plane measured from the velocity axis,
and $\mathbf R(\phi)$ the rotation matrix.

The port states are obtained by normalizing $\mathbf A$ to
$\mathbf T=\sqrt2\,\mathbf A/\|\mathbf A\|_F$, with $\|\cdot\|_F$ the
Frobenius norm. Both columns are scaled by the same factor, so the
signal level remains in $g$ and the ratio of the singular values,
$\cos\theta$ by~\eqref{eq:jsat_proj}, is preserved. The payload
factor thus has condition number $\kappa_{\mathrm{sat}}=1/\cos\theta$,
independent of $\phi$. A per-port
normalization, as for a large array matched at every scan
angle~\cite{hannan1964}, would lower $\kappa_{\mathrm{sat}}$ to
between 1 and $1/\cos\theta$ depending on $\phi$, so the common
scaling is the conservative choice.

\subsection{Faraday Rotation $\mathbf R_F$}
\label{sec:rf}
The ionosphere rotates the plane of polarization of the downlink by the
Faraday angle $\theta_F$. The factor $\mathbf R_F(t)$ is thus the
rotation by $\theta_F(t)$ about the propagation direction
$\hat{\mathbf k}$ from the satellite to the handset, with
\begin{equation}
	\theta_F=\frac{K}{f_c^2}\int_{\mathrm{UE}}^{\mathrm{sat}}
	N_e(\ell)\,\mathbf B_0(\ell)\cdot\hat{\mathbf k}\,\mathrm d\ell,
	\label{eq:faraday}
\end{equation}
where $\ell$ is the position along the ray, $N_e$ the electron density,
and $\mathbf B_0$ the geomagnetic flux density. The constant
$K\approx2.36\times10^{4}$ is in SI units~\cite[Sec.~4.3]{itu531}. The flux density is taken from the International Geomagnetic
Reference Field~\cite{beggan2026igrf14} and the electron density from
NeQuick~G~\cite{nequickg}. The ionization level of NeQuick~G is
scaled such that its vertical electron content matches a measured global ionosphere
map~\cite{upcgim2024} at the nadir point, as the
satellite lies below the shell height of the map, 450\,km.

\subsection{Environment Factor $\mathbf C_{\mathrm{env}}$}
\label{sec:cenv}
We consider a static handset in a suburban environment with a line of
sight to the satellite, i.e., the GOOD state of the LMS
channel~\cite{itu681}. In this state, the direct signal is received
together with multipath from buildings and trees. The
environment factor combines the dual-polarized channel structure
of~\cite{liolis2010}, the time-series synthesis of~\cite{itu681}, and
the S-band measurements of~\cite{burkhardt2014quadriga,kingthesis2007}. At pass snapshot $n$,
i.e., at time $t_n$ of~\eqref{eq:cascade}, $\mathbf C_{\mathrm{env}}(n)$
is assembled as
\begin{equation}
	\begin{split}
		\mathbf C_{\mathrm{env}}(n) \;=\;& \mathrm{diag}\big(r_1(n), r_2(n)\big)\,\mathbf R_{\mathrm{los}}(n)\\
		&+\; \sqrt{\Omega(n)}\,\mathbf V \odot \big(\mathbf Q_{\mathrm{rx}}\,\mathbf Z(n)\,\mathbf Q_{\mathrm{tx}}\big),
	\end{split}
	\label{eq:cenv}
\end{equation}
the matrix form of Loo's direct-plus-multipath model~\cite{loo1985}
with the structure of~\cite{liolis2010}. The first term represents the
direct signal, whose linear amplitudes in the two co-polar paths, from
each payload port to the matching handset port, are denoted by $r_1$
and $r_2$ and given in~\eqref{eq:adir}. Its deterministic
polarization geometry is captured by $\mathbf R_{\mathrm{los}}(n)$, the
rotation between the transverse port basis of the payload and the
receive basis of the handset along the LOS. The second term is the multipath signal, with $\Omega(n)$ the
multipath power relative to the unshadowed direct signal and $\odot$
the elementwise product. The multipath field $\mathbf Z(n)$ is multiplied
on both sides by $\mathbf Q_{\mathrm{tx}}$ and $\mathbf Q_{\mathrm{rx}}$,
the square roots of the $2\times2$ transmit and receive correlation
matrices with off-diagonal entries
$\rho_{\mathrm{tx}}=\rho_{\mathrm{rx}}=0.5$~\cite[Table~I]{liolis2010}.
The matrix $\mathbf V$ splits the multipath power between the co- and
cross-polar entries, with $\sqrt{1-\eta}$ on and $\sqrt{\eta}$ off the
diagonal, where $\eta=1/(1+10^{\mathrm{XPR}/10})$. Here, $\mathrm{XPR}$
is the ratio of co- to cross-polar multipath power in decibels, taken
per elevation from the S-band LMS parameters of QuaDRiGa for the
LOS state~\cite{burkhardt2014quadriga}.

The stochastic quantities in~\eqref{eq:cenv} follow the two-state
(GOOD/BAD) time-series synthesis of~\cite[Annex~1, Sec.~6.2]{itu681},
restricted to the GOOD state, with all parameters from its 1.5 to
3\,GHz table~\cite[Annex~2, Sec.~2.2]{itu681}. The synthesis
parameterizes its processes in meters traveled by a mobile terminal.
For a static handset, the equivalent walk
\begin{equation}
	\Delta d_{\mathrm w}(n) \;=\; D_{\mathrm{obs}}\,\Delta\Xi(n)
	\label{eq:walk}
\end{equation}
is the distance a mobile terminal would travel from snapshot $n{-}1$
to $n$. Here, $\Delta\Xi(n)$ is the angle between the two LOS
directions, and $D_{\mathrm{obs}}=25$\,m is the distance from
the handset to the obstacles. Along the walk, the synthesis
generates the mean direct level $m_A(n)$, its standard deviation
$\varsigma_A(n)$, and the multipath power $m_P(n)$, all in decibels,
so that $\Omega(n)=10^{m_P(n)/10}$. The walk is divided into events of
lognormal length, each with its own mean direct level around the
tabulated mean $\mu_A$, joined by linear transitions. The direct amplitudes are then
\begin{equation}
	\begin{gathered}
		(r_1, r_2)^{\!\top} = 10^{\,[m_A(n)\,\mathbf{1} + \varsigma_A(n)\,\mathbf Q_{\mathrm{dir}}\,\mathbf u(n)]/20},\\
		\mathbf Q_{\mathrm{dir}} = \sqrt{\left[\begin{smallmatrix}1&\rho_{\mathrm{dir}}\\
				\rho_{\mathrm{dir}}&1\end{smallmatrix}\right]},
	\end{gathered}
	\label{eq:adir}
\end{equation}
with $\mathbf 1=(1,1)^{\!\top}$, $\mathbf u(n)$ two unit-variance
Gauss-Markov processes with the tabulated correlation length $L_c$
along the walk, and $\rho_{\mathrm{dir}}=0.76$ the measured
correlation of the two direct amplitudes~\cite{kingthesis2007}. The
multipath field $\mathbf Z(n)$ has independent $\mathcal{CN}(0,1)$
entries, correlated between snapshots $n{-}1$ and $n$ by
$J_0(2\pi\,\Delta d_{\mathrm w}(n)/\lambda)$~\cite{clarke1968}, with
$J_0$ the zero-order Bessel function and $\lambda$ the carrier
wavelength.

The table of~\cite{itu681} gives its parameters in
elevation blocks from $20^\circ$ to $70^\circ$ and prescribes the
nearest block. We instead
interpolate each parameter linearly in elevation and keep the
$70^\circ$ block above $70^\circ$, so that its GOOD-state value
$\mu_A=-1.84$\,dB applies up to zenith.
The cross-polar ratio is kept constant above $40^\circ$, the highest
bin center of~\cite{burkhardt2014quadriga}. The
GOOD-state restriction, the walk~\eqref{eq:walk}, and the
interpolation are our additions to~\cite{itu681}.

\subsection{Handset Factor $\mathbf J_{\mathrm{UE}}$}
\label{sec:jue}
\begin{figure}[t]
	\centering
	\begin{tikzpicture}[baseline=(current bounding box.center), x={(0.43cm,0cm)}, y={(0.21cm,0.18cm)}, z={(0cm,0.43cm)},
		>=stealth, font=\scriptsize,
		enu/.style={->, black!45, thin},
		dev/.style={draw=black, fill=black!12, line width=0.4pt},
		bore/.style={->, very thick, red!75!black},
		lax/.style={->, thick, blue!70!black},
		lft/.style={->, thick, green!50!black}]
		\begin{scope}
			\draw[enu] (0,0,0) -- (1.3,0,0) node[right] {E};
			\draw[enu] (0,0,0) -- (0,1.5,0) node[above right=-1pt] {N};
			\draw[enu] (0,0,0) -- (0,0,1.5) node[above] {U};
			\draw[dev] (-0.8,-0.4,0) -- (0.8,-0.4,0) -- (0.8,0.4,0) -- (-0.8,0.4,0) -- cycle;
			\draw[lax] (0,0,0) -- (-1.2,0,0) node[left] {$z_b$};
			\draw[lft] (0,0,0) -- (0,1.2,0) node[right=1pt] {$y_b$};
			\draw[bore] (0,0,0) -- (0,0,1.15) node[left] {$x_b$};
			\node[right, align=left] at (2.1,0,0.2) {zenith\\$[0,\,90^\circ,\,0]$};
		\end{scope}
		\begin{scope}[yshift=-1.8cm]
			\draw[enu] (0,0,0) -- (1.3,0,0) node[above right=-1pt] {E};
			\draw[enu] (0,0,0) -- (0,1.5,0) node[above right=-1pt] {N};
			\draw[enu] (0,0,0) -- (0,0,1.5) node[above] {U};
			\draw[dev] (0,-0.4,-0.8) -- (0,0.4,-0.8) -- (0,0.4,0.8) -- (0,-0.4,0.8) -- cycle;
			\draw[lax] (0,0,0) -- (0,0,1.2) node[left] {$z_b$};
			\draw[lft] (0,0,0) -- (0,0.95,0) node[right=-1pt, yshift=-1pt] {$y_b$};
			\draw[bore] (0,0,0) -- (1.15,0,0) node[below=1pt] {$x_b$};
			\node[right, align=left] at (2.1,0,0.2) {fixed 0/0/0\\$[0,\,0,\,0]$};
		\end{scope}
		\begin{scope}[yshift=-3.6cm]
			\draw[enu] (0,0,0) -- (1.3,0,0) node[right] {E};
			\draw[enu] (0,0,0) -- (0,1.5,0) node[above right=-1pt] {N};
			\draw[enu] (0,0,0) -- (0,0,1.5) node[above] {U};
			\draw[dev] (-0.4,0,-0.8) -- (0.4,0,-0.8) -- (0.4,0,0.8) -- (-0.4,0,0.8) -- cycle;
			\draw[lax] (0,0,0) -- (0,0,1.2) node[left] {$z_b$};
			\draw[lft] (0,0,0) -- (-0.95,0,0) node[left] {$y_b$};
			\draw[bore] (0,0,0) -- (0,1.25,0) node[right] {$x_b$};
			\node[right, align=left] at (2.1,0,0.2) {fixed 0/0/90\\$[0,\,0,\,90^\circ]$};
		\end{scope}
	\end{tikzpicture}
	\begin{tikzpicture}[font=\scriptsize, >=stealth, baseline=(current bounding box.center)]
		\def\R{0.9}
		\foreach \yo/\ax/\axl/\tcol/\ncol in
		{0/green!50!black/y_b/green!50!black/blue!70!black,
			-2.75/blue!70!black/z_b/blue!70!black/green!50!black} {
			\begin{scope}[yshift=\yo cm]
				\foreach \g in {-10, 0} \draw[black!25] (0,0) circle ({(\g+20)/25*\R});
				\draw[black!25, dashed] (0,0) circle ({(5.3+20)/25*\R});
				\foreach \a in {0,30,...,330} \draw[black!12] (0,0) -- (\a:{25.3/25*\R});
				\draw[->, red!75!black, thick] (0,0) -- (0:{1.45*\R}) node[right] {$x_b$};
				\draw[->, \ax, thick] (0,0) -- (90:{1.4*\R}) node[right] {$\axl$};
				\draw[thick, black, smooth]
				plot[domain=-180:180, samples=181]
				(\x:{(5.3 - min(12*(\x/125)^2, 22.5) + 20)/25*\R});
				\draw[black!55, densely dashed] (0,0) -- (62.5:{22.3/25*\R});
				\draw[black!55, densely dashed] (0,0) -- (-62.5:{22.3/25*\R});
				\foreach \a in {0,60,-60,120,-120} {
					\pgfmathsetmacro\rc{(5.3 - min(12*(\a/125)^2, 22.5) + 20)/25*\R}
					\draw[->, \tcol, thick]
					([shift={(\a-90:0.09)}]\a:\rc+0.15) -- ([shift={(\a+90:0.09)}]\a:\rc+0.15);
					\node[\ncol, inner sep=0pt] at (\a:\rc-0.17) {$\otimes$};
				}
			\end{scope}
		}
	\end{tikzpicture}
	\caption{Handset attitudes $[\text{bank},\,\text{tilt},\,\text{heading}]$
		in the east/north/up frame on the left, with boresight $x_b$ (red),
		left axis $y_b$ (green), and long axis $z_b$ (blue). Element gain
		\eqref{eq:tr38901} of the handset antenna element in the $(x_b,y_b)$
		plane (top right) and the $(x_b,z_b)$ plane (bottom right), with
		rings at $-10$, $0$, and $5.3$\,dBi. The ports polarized along
		$\hat{\boldsymbol\theta}''$ (blue) and $\hat{\boldsymbol\phi}''$
		(green) are drawn as $\otimes$ where normal to the plane and as
		arrows where tangential.}
	\label{fig:attitude}
\end{figure}
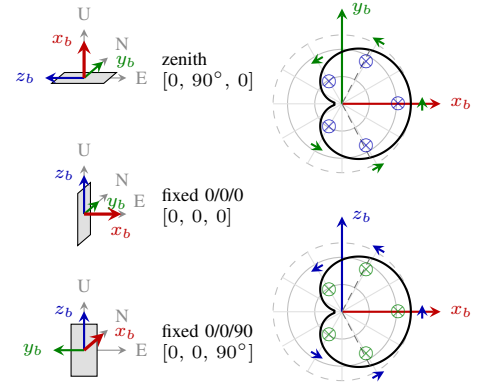

The handset is modeled as a flat panel without a chassis with
the standardized dual-polarized handset antenna
element~\cite[Table~7.3-2]{tr38901}, i.e., two co-located ports with
the common power pattern
\begin{equation}
	\begin{split}
		G_{\mathrm{el}}(\theta'',\phi'')&=5.3-\min\Big\{12\Big(\tfrac{\theta''-90^\circ}{125^\circ}\Big)^{2}\\
		&\qquad+12\Big(\tfrac{\phi''}{125^\circ}\Big)^{2},\,22.5\Big\}\ \mathrm{dBi}.
	\end{split}
	\label{eq:tr38901}
\end{equation}
The body frame of the handset, drawn in Fig.~\ref{fig:attitude}, has
the panel normal $x_b$ as boresight, $y_b$ pointing to the left, and
$z_b$ along the long axis. A direction $(\theta'',\phi'')$ has polar
angle $\theta''$ from $z_b$ and azimuth $\phi''$ from $x_b$ toward
$y_b$, so that the boresight lies at $\theta''=90^\circ$ and $\phi''=0$.
The first port is polarized along $\hat{\boldsymbol\theta}''$ and the
second along $\hat{\boldsymbol\phi}''$. The attitude
$[\text{bank},\text{tilt},\text{heading}]$ rotates the body frame
relative to the local east/north/up frame by
$\mathbf R_{\mathrm{att}}=\mathbf R_z(\text{heading})\,\mathbf R_y(-\text{tilt})\,
\mathbf R_x(\text{bank})$, with $\mathbf R_x$, $\mathbf R_y$, and
$\mathbf R_z$ the rotations about the east, north, and up axes. The
direction $-\hat{\mathbf k}$ to the satellite, expressed in the body
frame as $-\mathbf R_{\mathrm{att}}^{\top}\hat{\mathbf k}$, gives
$(\theta'',\phi'')$. By reciprocity, each row of
$\mathbf J_{\mathrm{UE}}$ projects the polarization vector of one port
onto the receive basis $(\hat{\boldsymbol\theta}_r,\hat{\boldsymbol\phi}_r)$.
These are the spherical unit vectors at the handset toward the
satellite, in which $\mathbf C_{\mathrm{env}}$ provides the field, so that
\begin{equation}
	\mathbf J_{\mathrm{UE}}=10^{G_{\mathrm{el}}/20}
	\begin{bmatrix}\hat{\boldsymbol\theta}''\!\cdot\hat{\boldsymbol\theta}_r &
		\hat{\boldsymbol\theta}''\!\cdot\hat{\boldsymbol\phi}_r\\
		\hat{\boldsymbol\phi}''\!\cdot\hat{\boldsymbol\theta}_r &
		\hat{\boldsymbol\phi}''\!\cdot\hat{\boldsymbol\phi}_r\end{bmatrix}.
	\label{eq:jue}
\end{equation}
Since both pairs of unit vectors are transverse to $\hat{\mathbf k}$,
the matrix in~\eqref{eq:jue} is orthogonal. The attitude therefore
acts on $\mathbf H$ only through the element gain and a rotation of
the polarization basis about the LOS. We simulate the three attitudes of Fig.~\ref{fig:attitude}, of which
the zenith attitude points the boresight up. The upright attitudes
$[\text{bank},\text{tilt},\text{heading}]=[0,0,0]^\circ$ and
$[0,0,90]^\circ$, referred to as fixed 0/0/0 and fixed 0/0/90, point it
east and north, respectively.

\section{Link Adaptation}
\label{sec:la}
The transmitter is assumed to know $\mathbf H$ perfectly and without
delay, so that the results depend on the channel alone. The SNR scales
with the slant range $d$ as
\begin{equation}
	\gamma=\gamma_0\,(d_0/d)^2,
	\label{eq:snr}
\end{equation}
where $d_0$ and $\gamma_0$ are the slant range and the reference SNR
at the nadir point, respectively. The reference SNR $\gamma_0$
in~\eqref{eq:snr} accounts for the transmit power, the free-space
path loss, and the receiver noise. The channel $\mathbf H$ contains
the antenna, polarization, and environment gains, so that all
attitudes and positions are compared at the same transmit power. The transmit power is normalized to one in total over
the two ports and the noise power to one per receive port. Ideal
precoding with
Gaussian input sends one layer by maximum-ratio transmission on the
dominant mode for rank~1 and two layers with waterfilling over both
modes for rank~2,
\begin{equation}
	\begin{split}
		\mathrm{SE}_1&=\log_2\!\big(1+\gamma\,\sigma_1^2\big),\\
		\mathrm{SE}_2&=\textstyle\sum_{i=1}^{2}\log_2\!\big(1+\gamma\,p_i\sigma_i^2\big),
	\end{split}
	\label{eq:se_ideal}
\end{equation}
where $p_i=[\nu-1/(\gamma\,\sigma_i^2)]^+$ is the power fraction of
mode $i$. The water level $\nu$ is set such that $p_1+p_2=1$. Since
waterfilling contains rank~1 as the case $p_2=0$, the selected
$\mathrm{SE}=\max(\mathrm{SE}_1,\mathrm{SE}_2)$ of~\eqref{eq:se_ideal}
uses rank~2 exactly
when $(\sigma_2^{-2}-\sigma_1^{-2})/\gamma<1$, a threshold on $\kappa$
and the SNR.

Closed-loop precoding uses the NR 2-port Type~I
codebook~\cite[Table~5.2.2.2.1-1]{ts38214r19}. For
rank~1, the precoder is the vector
$\mathbf w_q=[1,\;j^q]^{\top}/\sqrt2$, $q=0,\dots,3$, with the largest
SNR $\gamma\,\|\mathbf H\mathbf w_q\|^2$. For rank~2, the candidates
are the two matrices
$\mathbf W_0=\frac12\left[\begin{smallmatrix}1&1\\1&-1\end{smallmatrix}\right]$
and
$\mathbf W_1=\frac12\left[\begin{smallmatrix}1&1\\j&-j\end{smallmatrix}\right]$.
With precoder $\mathbf W$, the signal-to-interference-plus-noise ratio
(SINR) of layer $l$ after minimum mean square error detection is
\begin{equation}
	\mathrm{SINR}_l=\Big[\big(\mathbf I+\gamma\,\mathbf W^{\mathrm H}\mathbf H^{\mathrm H}
	\mathbf H\mathbf W\big)^{-1}\Big]_{ll}^{-1}-1,
	\label{eq:mmse_sinr}
\end{equation}
where $(\cdot)^{\mathrm H}$ is the conjugate transpose and $\mathbf I$
the identity matrix. Each
candidate is scored by the bit-interleaved coded modulation (BICM)
mutual information~\cite{caire1998bicm}
$\sum_l C_{\mathcal M}(\mathrm{SINR}_l)$ per channel use, with
$\mathrm{SINR}_l$ from~\eqref{eq:mmse_sinr}. Here, $C_{\mathcal M}$ is
the per-symbol BICM mutual information of the constellation
$\mathcal M$, which is quadrature phase-shift keying (QPSK) or
16-point quadrature amplitude modulation (16QAM), as higher-order
modulations are not considered. Each candidate uses one modulation on both layers, as in one NR
codeword, choosing the modulation with the larger score. The rank and
precoder with the highest score are selected, with rank~1 in case of a
tie. The resulting SE accounts for the finite alphabet but not for
coding or block errors. The rank-1 reference of
Section~\ref{sec:sim} is the best rank-1 candidate, or
$\mathrm{SE}_1$ for ideal precoding.

\section{Simulation Results}
\label{sec:sim}
\subsection{Simulation Setup}
\label{sec:setup}
We consider a satellite in a circular orbit 330\,km above the
equatorial Earth radius (342\,km above the nadir point) with
$53^\circ$ inclination, equipped with the payload of
Section~\ref{sec:jsat} and transmitting at 2\,GHz. The simulations evaluate the satellite at the
single position above Munich ($48.1^\circ$\,N, $11.6^\circ$\,E), the
nadir point of the maps. Handsets are placed on rings of 10\,km spacing
around the nadir point, giving 7390 positions out to 480\,km. The
footprint is assumed to be limited to the area with elevation larger
than $32^\circ$. The environment is the suburban GOOD state of
Section~\ref{sec:cenv}, and the amplitude $a(t)$ in~\eqref{eq:cascade}
contains ionospheric scintillation with the index
$S_4=0.3$~\cite{itu531}. The Faraday angle is taken from the global ionosphere
map~\cite{upcgim2024} of a day with a geomagnetic storm of the highest
level, G5~\cite{noaa2024g5}. Since the matrix in~\eqref{eq:jue} is orthogonal, this
rotation affects the results only through the non-orthogonal part of
$\mathbf C_{\mathrm{env}}$. Each position is evaluated for $10^4$
independent realizations drawn from the stationary distributions of
Section~\ref{sec:cenv}, so the walk~\eqref{eq:walk} does not affect
the maps. The standard error of a rank-2
probability is then below 0.005 per position, which moves the 0.95
contours by less than 4\,km.

In the zenith attitude, a handset at the nadir point receives about
$\gamma_0+3.7$\,dB summed over its two ports, namely the element gain
of 5.3\,dBi and the mean GOOD-state power of $-1.6$\,dB. The latter
is $\mu_A=-1.84$\,dB plus the direct-level spread and the multipath.
For comparison, the NR~NTN calibration case
SC10~\cite[Table~6.1.1.2-1]{tr38821} has a median SNR near 16\,dB
without interference.

\subsection{Rank-2 Areas}

\begin{figure}[t]
	\centering
	\input{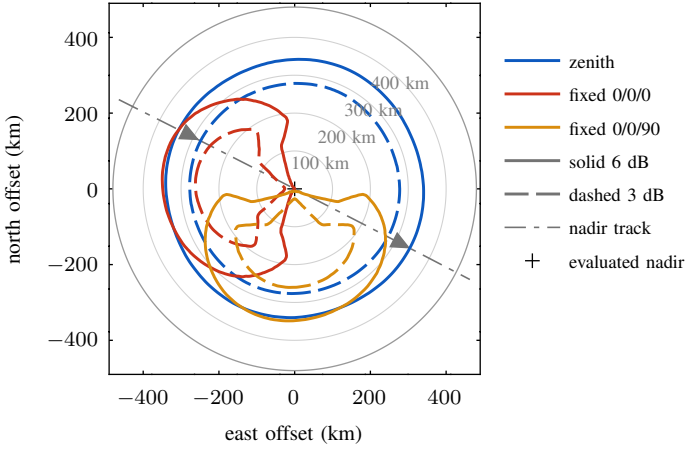}
	\caption{Areas in which ideal precoding selects rank~2 with a
		probability of at least 0.95 at $\gamma_0=6$\,dB and 3\,dB.}
	\label{fig:contours}
\end{figure}

In Fig.~\ref{fig:contours}, the rank-2 area of the zenith attitude is
a circle of about 340\,km radius at 6\,dB and 280\,km at 3\,dB. It is round because
$\kappa_{\mathrm{sat}}=1/\cos\theta$ of~\eqref{eq:jsat_proj} does not
depend on $\phi$. The upright attitudes select rank~2 only on one side
of the nadir point, west of it for fixed 0/0/0 and south of it for
fixed 0/0/90. Handsets on the other side have the satellite behind
them, where the element gain is lower by up to 22.5\,dB, the
front-to-back ratio of~\eqref{eq:tr38901}, and near the nadir point
the LOS runs along $z_b$, where the gain is also low. Since $\mathbf J_{\mathrm{UE}}$
in~\eqref{eq:jue} is orthogonal, these areas are determined by the
element gain alone.

\subsection{Spectral Efficiency}

\begin{table}[t]
	\caption{Mean SE over the footprint in bits per channel use and gain
of rank selection over rank~1 for ideal precoding with Gaussian input
and for closed-loop precoding with QPSK/16QAM}
	\label{tab:se}
	\centering
	\footnotesize
	\setlength{\tabcolsep}{3.5pt}
\begin{tabular}{@{}rlcccccc@{}}
\toprule
& & \multicolumn{3}{c}{ideal precoding} &
\multicolumn{3}{c}{closed-loop precoding}\\
\cmidrule(lr){3-5}\cmidrule(l){6-8}
$\gamma_0$ & attitude & rank 1 & selection & gain & rank 1 & selection & gain\\
\midrule
13\,dB & zenith & 4.04 & 5.53 & 37\,\% & 3.32 & 4.94 & 49\,\%\\
 & fixed 0/0/0 & 2.14 & 2.67 & 25\,\% & 1.87 & 2.44 & 31\,\%\\
 & fixed 0/0/90 & 2.19 & 2.73 & 25\,\% & 1.92 & 2.50 & 31\,\%\\
\midrule
6\,dB & zenith & 2.09 & 2.46 & 18\,\% & 1.86 & 2.26 & 22\,\%\\
 & fixed 0/0/0 & 0.92 & 1.00 & 8\,\% & 0.83 & 0.93 & 12\,\%\\
 & fixed 0/0/90 & 0.95 & 1.03 & 8\,\% & 0.85 & 0.95 & 12\,\%\\
\midrule
3\,dB & zenith & 1.43 & 1.59 & 11\,\% & 1.28 & 1.47 & 15\,\%\\
 & fixed 0/0/0 & 0.58 & 0.61 & 4\,\% & 0.53 & 0.57 & 7\,\%\\
 & fixed 0/0/90 & 0.60 & 0.63 & 4\,\% & 0.54 & 0.58 & 7\,\%\\
\bottomrule
\end{tabular}

\end{table}

\begin{figure}[t]
	\centering
	\input{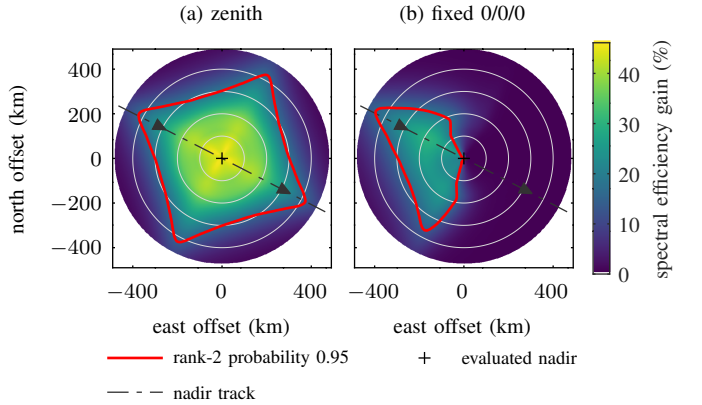}
	\caption{SE gain of rank selection over rank~1 with closed-loop
		precoding at $\gamma_0=6$\,dB for the zenith attitude (a) and fixed
		0/0/0 (b), with the contour of rank-2 probability 0.95.}
	\label{fig:gainmaps}
\end{figure}

Table~\ref{tab:se} lists the mean SE over the footprint for both
schemes. At 13\,dB, rank selection raises the mean SE of the zenith
attitude by 37\,\% with ideal precoding and by 49\,\% with closed-loop
precoding. For the upright attitudes, the
gains are about 25\,\% and 31\,\%. The relative gain is larger with
closed-loop precoding for two reasons. At 13\,dB, rank~1 cannot
exceed the 4\,bits per channel use of 16QAM, whereas rank~2 can. At
every SNR, the codebook reduces the SE of rank~1 more than that of rank~2. A rank-1
precoder $\mathbf w_q$ loses up to 3\,dB when the dominant right
singular vector of $\mathbf H$ coincides with one payload port,
whereas the rank-2 matrices are unitary up to a scale factor. Fig.~\ref{fig:gainmaps} maps the
closed-loop gain at 6\,dB, which peaks at 46\,\% near the nadir
point for the zenith attitude and at 30\,\% 180\,km
west-northwest of it for fixed 0/0/0. The closed-loop rank-2 area of
the zenith attitude is a square with its corners along and across the
nadir track. There, the payload ports are orthogonal and none of the
$\mathbf w_q$ matches the dominant port, so that rank~2 is selected
at larger distances than with ideal precoding.

\subsection{Percentage of Satellite Pass in Rank~2}
\label{sec:pass}
\begin{figure}[t]
	\centering
	\input{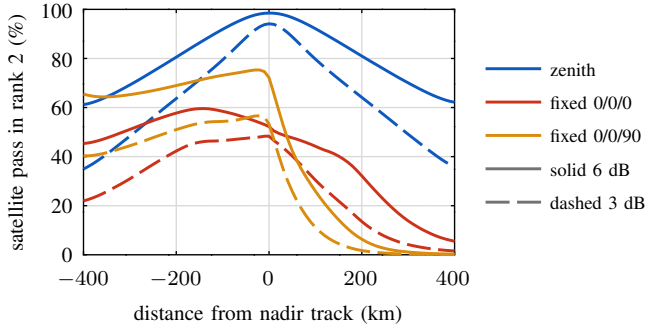}
	\caption{Percentage of the satellite pass above $32^\circ$ elevation
		in rank~2 with closed-loop precoding versus the signed distance from
		the nadir track.}
	\label{fig:pass}
\end{figure}

For a handset at a signed distance from the nadir track, positive to
the left of the flight direction, the pass is a line through the maps parallel to the track. This neglects
the change of the Faraday angle along the track.
Fig.~\ref{fig:pass} shows the percentage of the pass in rank~2, i.e.,
the mean closed-loop rank-2 probability along this line.
A zenith-attitude handset on the nadir track is in rank~2 for 99\,\%
of the pass at 6\,dB and 94\,\% at 3\,dB, and 200\,km from the track
for 81\,\% and 64\,\%. At 6\,dB, the upright attitudes peak at
60\,\% for fixed 0/0/0, 140\,km to the right of the track, and at
75\,\% for fixed 0/0/90, 30\,km to the right. They fall to 32\,\%
and 6\,\% at 200\,km to the left, where the satellite is behind the
handset.

\section{Conclusions}
\label{sec:concl}
The advantage of a second polarization layer on the downlink from an
S-band LEO satellite to a handset has been evaluated over the
footprint. For a handset in the zenith attitude, ideal precoding
selects rank~2 with a probability of at least 0.95
within about 340\,km of the nadir point at a reference SNR of 6\,dB.
Beyond that distance, the condition number and scan loss of the
payload and the path loss reduce the benefit of a second layer. Over
the footprint, rank selection with closed-loop precoding raises the
mean SE by 49\,\% for this handset and by 31\,\% for upright handsets
at 13\,dB, and by 15\,\% and 7\,\% at 3\,dB. The relative gain is
larger with the NR codebook than with ideal precoding. The rank-1
codebook loses up to 3\,dB where the dominant mode is a single payload
port, and 16QAM limits rank~1 at 13\,dB. A zenith-attitude
handset on the nadir track uses rank~2 during 99\,\% of a pass at
6\,dB. For upright handsets, the element gain restricts rank~2 to
the part of the footprint in which the handset faces the satellite, so
that the attitude has a larger effect than the polarization geometry.
A single satellite can thus provide a second polarization layer to a
handset with the existing NR codebook over a large part of its
footprint, provided that the handset faces the satellite. These
results hold for ideal cross-polarized handset ports, perfect
channel knowledge without delay, the GOOD state, a single orbit
height, and a modulation alphabet that stops at 16QAM, which limits rank~1 at 13\,dB. Port isolation of real handsets, feedback delay, shadowed
states, and block errors are left for future work.

\balance
\bibliographystyle{IEEEtran}
\bibliography{eucap}

\end{document}